\documentclass[final,3p,times,fleqn]{elsarticle}

\pdfoutput=1

\usepackage{amssymb}
\usepackage{amsmath}
\usepackage{amsthm}
\usepackage{graphicx}

\usepackage{lineno}

\makeatletter
\def\ps@pprintTitle{%
 \let\@oddhead\@empty
 \let\@evenhead\@empty
 \def\@oddfoot{\footnotesize\itshape
 \ifx\@journal\@empty
 \else\@journal\fi\hfill\today}%
 \let\@evenfoot\@oddfoot}
\makeatother

\begin{document}

\begin{frontmatter}


\begin{center}
Note
\end{center}

\title{Wrinkle-Free Interface Compression for Two-Fluid Flows}

\author[ad1]{Yashar Mehmani\corref{cor1}}
\address[ad1]{Department of Energy Resources Engineering, Stanford University, 367 Panama St., Green Earth Sciences 050, Stanford, CA 94305-4007, USA}
\cortext[cor1]{Corresponding author: Yashar Mehmani}
\ead{ymehmani@stanford.edu}

\begin{abstract}

Pure advection of a conservative scalar is relevant to several applications including two-phase flow. Successful numerical schemes must capture the sharp interface between the phases while maintaining a smooth (wrinkle-free) interfacial contour. Among other methods, algebraic schemes that are based on the addition of an extra artificial interface compression term to the scalar transport equation have gained recent popularity (e.g., OpenFOAM; porous media). While the approach offers unique flexibilities in spatiotemporal discretizations, common numerical treatments result in the wrinkling of the interface. In this short note, an easy-to-implement solution to this problem is presented.
\end{abstract}


\end{frontmatter}

\biboptions{numbers,sort&compress}

\section{Problem statement}
\label{sec:prob}
Consider Eq. \ref{eq:regtrans} for the transport of a scalar $\alpha$, bounded between $0$ and $1$, under the solenoidal velocity field $\vec{u}$.
	\begin{equation} \label{eq:regtrans}
	\partial_{t}\,\alpha + \nabla.\left(\vec{u}\,\alpha\right) = 0
	\end{equation}

Eq. \ref{eq:regtrans} is solved in two-phase flow simulations to describe the motion of each phase represented by regions where $\alpha=0$ or $1$ (Fig. \ref{fig:schem}a). High-quality solutions of Eq. \ref{eq:regtrans} are necessary to avoid so-called ``spurious currents'' at the two-phase interface represented by the $\alpha\approx0.5$ contour \cite{scardovelli1999direct}. Specifically, the transition between $\alpha=0$ to $1$ must remain sharp and free of wrinkles. It is known that even high-order discretizations of $\nabla.(u\alpha)$ cause unacceptable levels of numerical diffusion \cite{hirt1981VOF}. Two main classes of compressive schemes have thus been developed. The first is based on local geometric reconstructions of the interface, such as PLIC \cite{debar1974PLIC,youngs1982PLIC}, which is among the most accurate but computationally complex and therefore difficult to implement. The second is based on algebraic discretizations of Eq. \ref{eq:regtrans}, which is the focus herein. Examples of algebraic schemes include CICSAM \cite{ubbink1999CICSAM}, HRIC \cite{muzaferija1999HRIC}, HiRAC \cite{heyns2013HiRAC}, STACS \cite{darwish2006STACS}, and THINC \cite{xie2017THINC}. Further details can be found in the recent review \cite{mirjalili2017review}. CICSAM is one of the most successful, which has led to further improvements by \cite{waclawczyk2008MCICSAM, zhang2014MCICSAM}. All schemes, except THINC, discretize $\nabla.(u\alpha)$ via a weighted sum of a compressive downwind scheme and a high-resolution upwind scheme \cite{patel2015framework}. The weight is a function of the angle between the interface normal $\vec{n}_i=\nabla\alpha/||\nabla\alpha||$ and the cell face normal $\vec{n}_f$. This weighting ensures the interface remains compact when $\vec{n}_i$ is parallel to $\vec{n}_f$ and wrinkle-free when $\vec{n}_i$ is normal to $\vec{n}_f$.

An alternative to compressive discretizations of $\nabla.(u\alpha)$ is adding a separate interface compression term to Eq. \ref{eq:regtrans} as shown in Eq. \ref{eq:cmptrans} \cite{harten1974intcomp, rusche2003intcomp}. $\vec{u}_r$ is a user-defined velocity along $\vec{n}_i$. Eq. \ref{eq:cmptrans} decouples advection from interface compression providing some flexibility and simplicity in spatial and temporal descretizations (implicit/explicit). However, common treatments of the compressive term in Eq. \ref{eq:cmptrans}, e.g., OpenFOAM \cite{openfoam2018}, cause the interface to wrinkle when $\vec{n}_i$ is normal to $\vec{n}_f$. Section \ref{sec:disc} presents a simple discretization of Eq. \ref{eq:cmptrans} that preserves a sharp yet smooth interface.
	\begin{equation} \label{eq:cmptrans}
		\partial_{t}\,\alpha + \nabla.\left(\vec{u}\,\alpha\right) + \nabla.\left(\alpha(1-\alpha)\,\vec{u_r}\right) = 0
	\end{equation}

\section{Discretization}
\label{sec:disc}
Eq. \ref{eq:FV} is the finite-volume representation of Eq. \ref{eq:cmptrans} for grid cell $c$ comprised of faces $f$. $F^a_f$ and $F^c_f$ are the advective and compressive face fluxes,  respectively. $A_f$ denotes the face area. Summation is over all faces of $c$. $\alpha$ is understood here as a cell-averaged quantity. $F^a_f$ and $F^c_f$ are discretized separately with a second-order Godunov upwind scheme \cite{leveque2002hyper}, biased against their characteristic wave velocities, $\vec{u}_w$, in Eq. \ref{eq:cmptrans}. If we rewrite Eq. \ref{eq:cmptrans} as Eq. \ref{eq:char} (note $\nabla.u=0$), we see that $\vec{u}_w=\vec{u}$ and $\vec{u}_w=\vec{u}_c$ (annotated) are the advective (for $F^a_f$) and compressive (for $F^c_f$) wave velocities, respectively. The rightmost term in Eq. \ref{eq:char} is a source term proportional to interfacial curvature $\kappa$ (note $u_r \varpropto \vec{n}_i$).
	\begin{align}
	&\partial_{t}\,\alpha + \underset{\forall f}{\sum} F^a_f A_f +
	                        \underset{\forall f}{\sum} F^c_f A_f = 0 \label{eq:FV} \\
	&\partial_{t}\,\alpha + \vec{u}.\,\nabla\alpha + 
	                        \underset{=\vec{u}_c}{\underbrace{(1-2\alpha)\,\vec{u}_r}}.\nabla\alpha + 
	                        \alpha(1-\alpha)\,\underset{\varpropto\, \kappa=\nabla.\vec{n}_i}
	                                   {\underbrace{\nabla.\vec{u}_r}} = 0  \label{eq:char}
	\end{align}

$F^a_f$ and $F^c_f$ are computed from Eq. \ref{eq:advdisc}-\ref{eq:cmpdisc}. Subscript $f$ denotes quantities defined at cell faces. Superscript $\downarrow$ denotes upwind-biased values with respect to  $\vec{u}_w$. $\vec{n}_f$ is the face normal. In Eq. \ref{eq:cmpdisc}, we used $\vec{u}_r=||u_r||\,\vec{n}_i$. It is common to compute $||u_r||_f$ in Eq. \ref{eq:cmpdisc} from Eq. \ref{eq:cnstcmp} \cite{rusche2003intcomp, openfoam2018}, where $\Lambda_f=1$ and $\zeta$ is fixed between $1$ and $2$. The \textit{max} operator acts over all faces in the domain. Eq. \ref{eq:cnstcmp} implies compression rates proportional to the face flux magnitude, i.e., $|\vec{u}.\vec{n}_f|_f$.
	\begin{align}
	&F^a_f = \left(\alpha\vec{u}\right)_f.\vec{n}_f \approx 
	         \alpha^{\downarrow}_f \left(\vec{u}.\vec{n}_f\right)_f \label{eq:advdisc} \\
	&F^c_f = \left(\alpha(1-\alpha) ||u_r|| \vec{n}_i \right)_f . \vec{n}_f \approx
             \left(\alpha(1-\alpha)\right)^{\downarrow}_f ||u_r||_f 
                                  \left(\vec{n}_i.\vec{n}_f\right)_f \label{eq:cmpdisc} \\
	&||u_r||_f = \Lambda_f\min\left[\zeta\,|\vec{u}.\vec{n}_f|_f, \underset{\forall f_i}{\max} 
	                         |\vec{u}.\vec{n}_{f_i}|_{f_i} \right] \label{eq:cnstcmp}
	\end{align}
	
	\begin{figure} \vspace*{-1.5cm}
	\centerline{\includegraphics[scale=0.6,trim={2.5cm 27cm 5cm 3.4cm},clip]{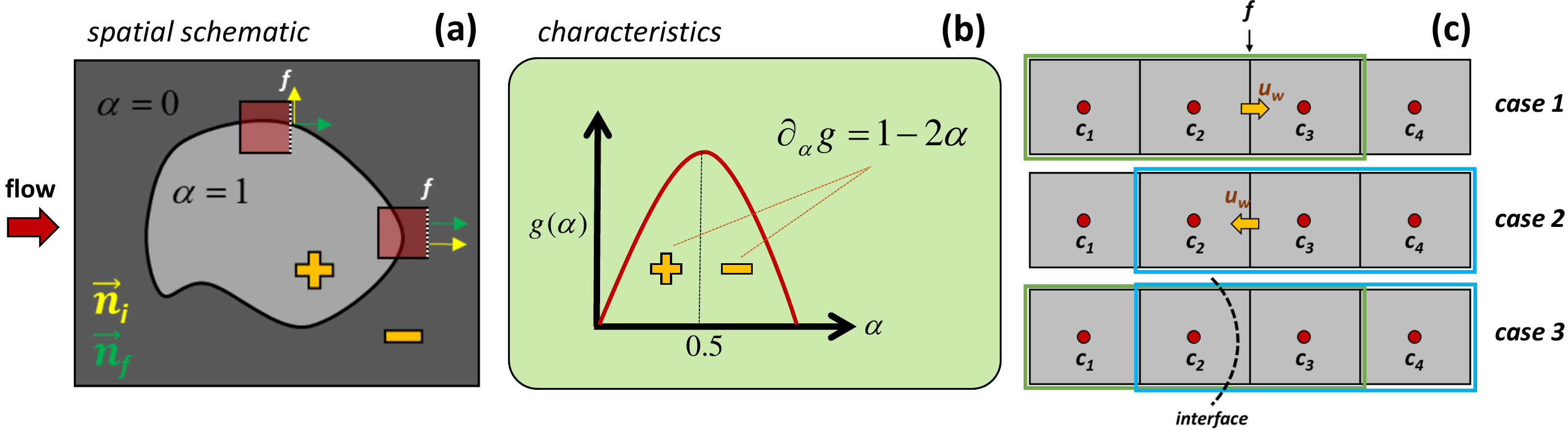}}
	\caption{(a) Cartoon of the spatial distribution of $\alpha$. Normals to the interface, $\vec{n}_i$ (yellow), and face $f$ (white), $\vec{n}_f$ (green), are shown for two cells (red). (b) The sign of the derivative of $g(\alpha)$ determines the direction of the compressive wave velocity $\vec{u}_c$ in Eq. \ref{eq:char}. (c) Stencils used to compute $F^a_f$ and $F^c_f$ at $f$ are shown, where $c_i$ denotes cells. Case 1 shows the stencil (green box) used when $\vec{u}_w$ points from left-to-right. Case 2 shows the stencil  (blue box) used when $\vec{u}_w$ points from right-to-left. Case 3 is considered only in computing $F^c_f$ (Eq. \ref{eq:galph}) and corresponds to when an interface exists between $c_2$ and $c_3$ (unlike Cases 1-2). For $F^a_f$, the presence of an interface between $c_2$ and $c_3$ is immaterial and only Cases 1-2 are considered.}
	\centering
	\label{fig:schem}
	\end{figure}
	
To compute $\alpha^{\downarrow}_f$ and $(\alpha(1-\alpha))^{\downarrow}_f$, we limit our discussion to Cartesian grids. While extensions of what follows to unstructured grids is conceivable, they were not tested. Without loss of generality, consider the 1D stencil around face $f$ in Fig. \ref{fig:schem}c. Treatments of 2D/3D follow by straighforward generalizations of the 1D case. Define $g(\alpha)=\alpha(1-\alpha)$ as shown in Fig. \ref{fig:schem}b. The upwind direction is biased against $\vec{u}_w=\vec{u}$ for $F^a_f$ and $\vec{u}_w=\vec{u}_c$ for $F^c_f$. Note that $\vec{u}_c$ always points towards the interface (Figs. \ref{fig:schem}a-b). Fig. \ref{fig:schem}c depicts the upwind stencils for when $\vec{u}_w$ points from left-to-right (Case 1) and right-to-left (Case 2). For $F^a_f$, set $\vec{u}_w=\vec{u}$ and compute $\alpha^{\downarrow}_f$ via Eq. \ref{eq:lim}, in which $\alpha_{I}=\alpha_{c_1}$, $\alpha_{II}=\alpha_{c_2}$, $\alpha_{III}=\alpha_{c_3}$ for Case 1, and $\alpha_{I}=\alpha_{c_4}$, $\alpha_{II}=\alpha_{c_3}$, $\alpha_{III}=\alpha_{c_2}$ for Case 2. Eq. \ref{eq:lim} is the flux-limited 2\textsuperscript{nd}-order QUICK scheme \cite{leonard1988QUICK}. For $F^c_f$, set $\vec{u}_w=\vec{u}_c$ and compute $(\alpha(1-\alpha))^{\downarrow}_f$ via Eq. \ref{eq:galph}, in which $\alpha^{\downarrow}_f$ is obtained from Eq. \ref{eq:lim}. In Eq. \ref{eq:galph}, Case 3 corresponds to when the interface ($\alpha=0.5$) is between $c_2$ and $c_3$ (Fig. \ref{fig:schem}c), whereas Cases 1-2 correspond to when no interface is between $c_2$ and $c_3$. Eq. \ref{eq:galph} is the 2\textsuperscript{nd}-order Godunov scheme \cite{leveque2002hyper}. Finally, we propose Eq. \ref{eq:angcmp} for $\Lambda_f$ in Eq. \ref{eq:cnstcmp} (instead of $\Lambda_f=1$), where we set $\beta=1$ and $\vec{n}_i=\nabla\alpha/||\nabla\alpha||$. In this work, $\alpha$ is not smoothed when computing $\vec{n}_i$.
	\begin{subequations} \label{eq:lim}
	\begin{align}
	&\alpha^{\downarrow}_f = \alpha_{II} + \frac{1}{2}\psi(r) \left(\alpha_{III}-\alpha_{II}\right) \label{eq:lim1} \\
	&\psi(r) = \max\left[0,\min\left(2r,(3+r)/4,2\right)\right] \label{eq:lim2} \\
	&r = \frac{\alpha_{II}-\alpha_{I}}{\alpha_{III}-\alpha_{II}} \label{eq:lim3}
	\end{align}
	\end{subequations}
	\begin{equation} \label{eq:galph}
	\left[\alpha(1-\alpha)\right]^{\downarrow}_f = 
		\left\{\begin{array}{cc}
		g^+ = g(\alpha^{\downarrow}_f), \qquad \text{Case 1} \\[1pt]
		g^- = g(\alpha^{\downarrow}_f), \qquad \text{Case 2} \\[2pt]
		\min(g^+,g^-), \qquad \text{Case 3}
		\end{array}\right.
	\end{equation}
	
	\begin{equation} \label{eq:angcmp}
	\Lambda_f = \min \left\{\beta\frac{\cos(2\theta_f)+1}{2},1\right\}, \qquad
	\theta_f=\arccos|\vec{n}_i.\vec{n}_f|_f
	\end{equation}

Eq. \ref{eq:angcmp} scales the compression rate $||u_r||_f$ in Eq. \ref{eq:cnstcmp} in proportion to the angle between $\vec{n}_i$ and $\vec{n}_f$. Compression is largest when $\vec{n}_i$ and $\vec{n}_f$ are parallel and zero when perpendicular (Fig. \ref{fig:schem}a). Similar expressions for $\Lambda_f$ have been proposed by \cite{lee2015adaptive, piro2013adaptive, aboukhedr2018adaptive}. In \cite{piro2013adaptive}, the interface is compressed only when $\vec{n}_i.\vec{u}>0$ causing asymmetric distortions (\ref{appA}). In \citep{lee2015adaptive}, $\Lambda_f$ contains an explicit dependence on time step, which can cause CFL restrictions. Tests also showed that the expression for $\Lambda_f$ therein to be less accurate than Eq. \ref{eq:angcmp} (\ref{appA}). In \citep{aboukhedr2018adaptive}, $\Lambda_f$ neglects any dependence on $\vec{n}_f$ and adjusts the compression rate in proportion to the local interface thickness. Arguments in \cite{patel2015framework} suggest that such dependence to be key for interface capturing. Eq. \ref{eq:FV}-\ref{eq:angcmp} were solved for $\alpha$ both explicitly and semi-implicitly in time. The latter implies that $\vec{n}_i$ from the previous time step was taken as the only explicit quantity.

\section{Results}
\label{sec:result}
\begin{figure} [ht]
	\centerline{\includegraphics[scale=0.5,trim={0.3cm 8.6cm 0.2cm 3cm},clip]{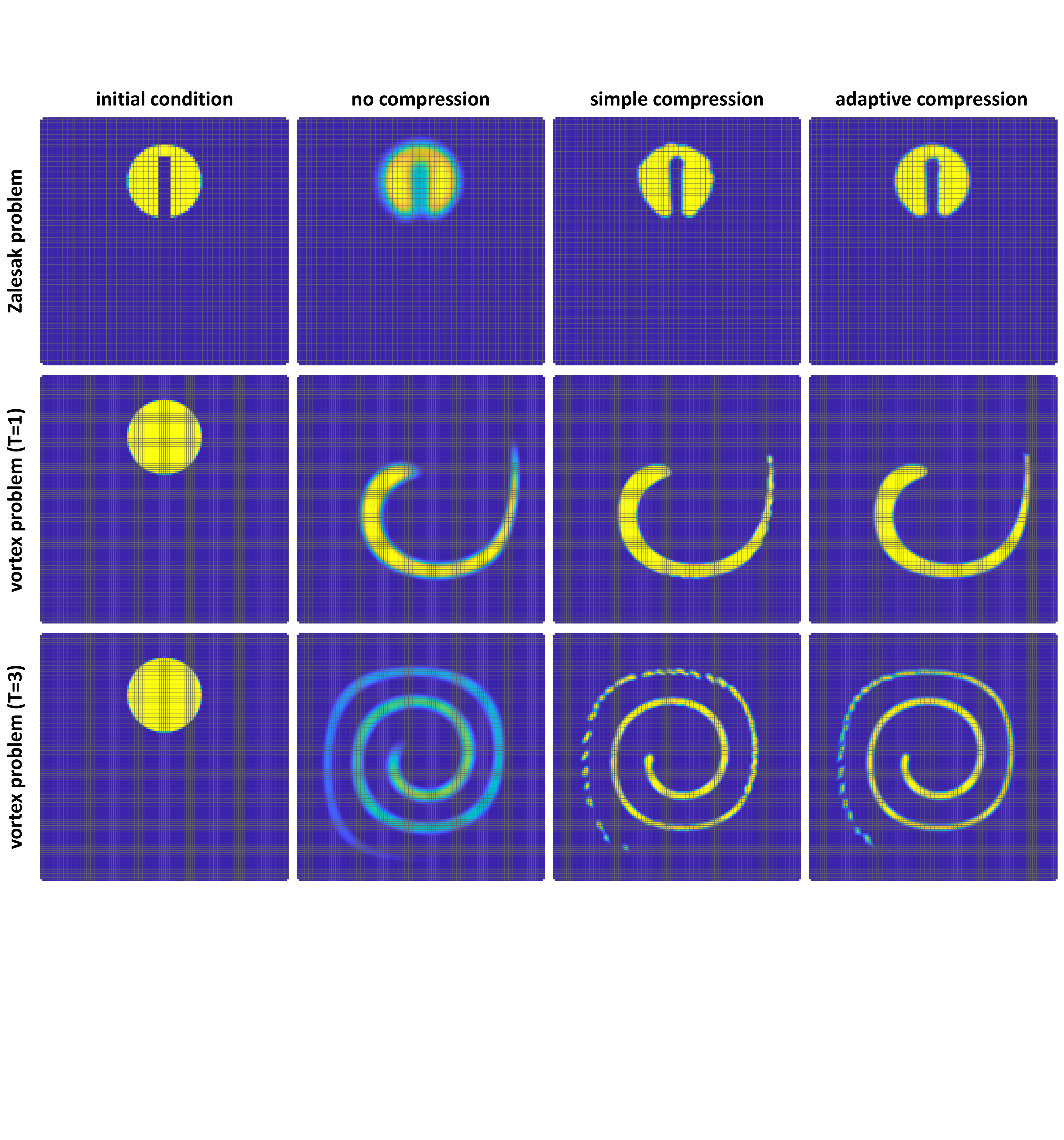}}
	\caption{Comparison between no compression ($\Lambda_f=0$), simple compression ($\Lambda_f=1$), and adaptive compression ($\Lambda_f$ from Eq. \ref{eq:angcmp}). (row 1) Zalesak's \cite{zalesak1979test} solid body rotation (one revolution). (row 2) Rider and Kothe's \cite{rider1998test} vortex shear rotation for $T=1$ and (row 3) for $T=3$.}
	\centering
	\label{fig:tests}
	\end{figure}

\begin{figure} [ht] \vspace*{-1.5cm}
	\centerline{\includegraphics[scale=0.55,trim={3.1cm 11.8cm 3.5cm 2cm},clip]{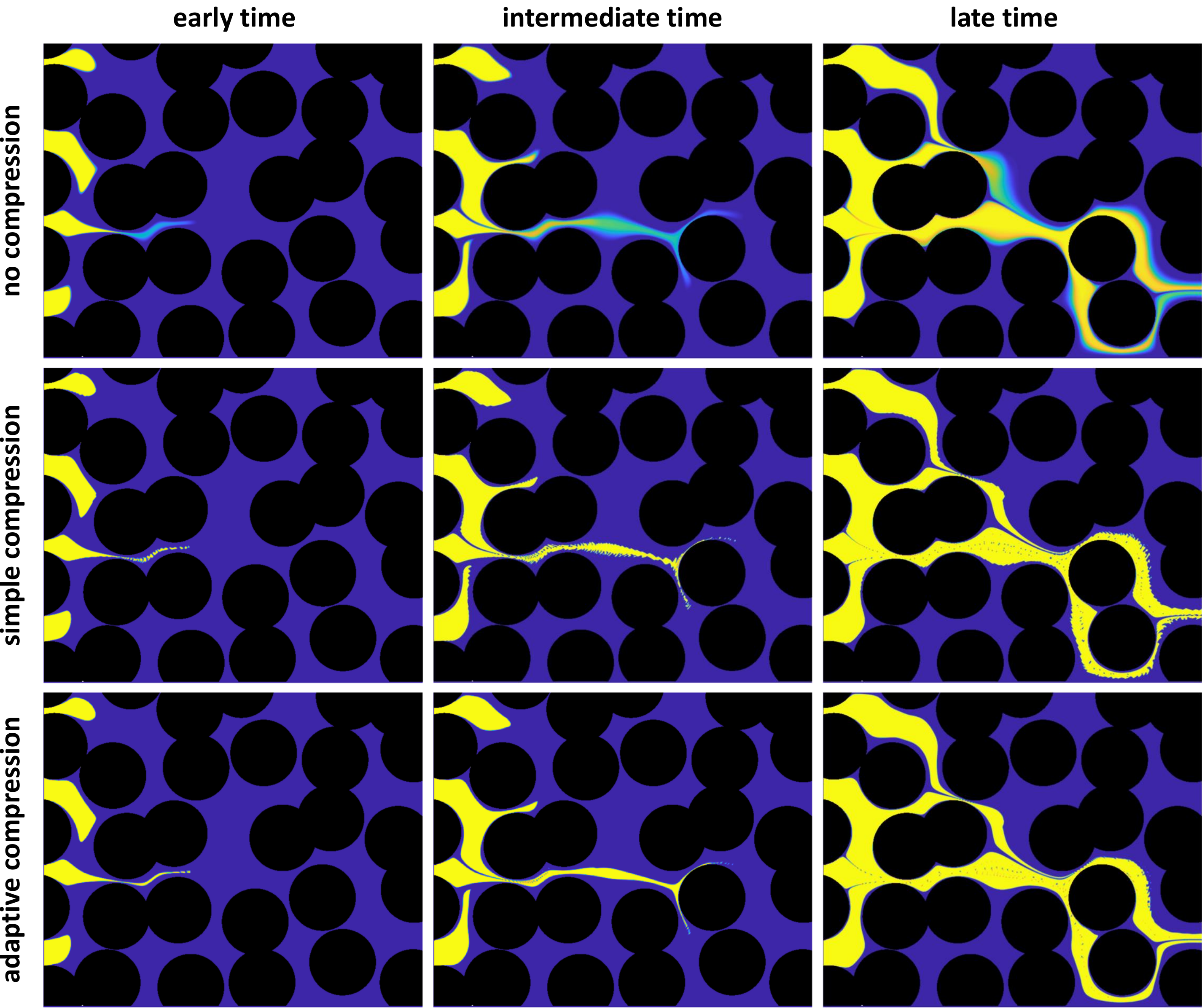}}
	\caption{Comparison between no compression ($\Lambda_f=0$), simple compression ($\Lambda_f=1$), and adaptive compression ($\Lambda_f$ from Eq. \ref{eq:angcmp}) for transport in a 2D disk pack at early, intermediate, and late times. Flow is from left to right and $\vec{u}$ is the solution to the Stokes equation.}
	\centering
	\label{fig:pores}
	\end{figure}

Three compression schemes were tested: (a) no compression with $\Lambda_f=0$, (b) simple compression with $\Lambda_f=1$, and (c) adaptive compression with $\Lambda_f$ in Eq. \ref{eq:angcmp}. Three test problems were considered: (test 1) Zalesak's \cite{zalesak1979test} solid body rotation, where the initial key-hole geometry in Fig. \ref{fig:tests} is rotated once counter-clockwise under $\vec{u}$ given by Eq. \ref{eq:zalesak}, (test 2) Rider and Kothe's \cite{rider1998test} vortical shear rotation, where the circle in Fig. \ref{fig:tests} is exponentially stretched up to times $T=1$ and $T=3$ under $\vec{u}$ given by Eq. \ref{eq:vortex}, and (test 3) flow in a porous disk pack (Fig. \ref{fig:pores}), where flow is from left-to-right and $\vec{u}$ is the solution to the Stokes equation. In test 3, streamlines split and merge in addition to simple shear and rotation. Domains in tests 1-2 are $1\times1$ squares with $100\times100$ cells. In Eq. \ref{eq:zalesak}-\ref{eq:vortex}, $x$ and $y$ denote horizontal and vertical directions and $x_0=y_0=0.5$. Figs. \ref{fig:tests}-\ref{fig:pores} show that adaptive compression is the most accurate among the schemes tested. $\Lambda_f=0$ expectedly results in high levels of numerical diffusion (despite 2\textsuperscript{nd}-order). $\Lambda_f=1$ causes break-up of thin filaments and the wrinkling of the interface, where $\vec{n}_i$ is normal to $\vec{u}$ (improvements are minor for other constant $\Lambda_f$). Such interfacial distortions are known to lead to parasitic currents near the interface in capillary-dominated two-phase flow simulations \cite{scardovelli1999direct}, which are important in porous media applications. The adaptive compression scheme of Section \ref{sec:disc} thus provides a simple yet accurate method for capturing sharp interfaces.
	\begin{align}
	&\vec{u} = \left(y_0-y\;,\, x-x_0\right) \label{eq:zalesak} \\
	&\vec{u} = \left(-\sin^2(\pi x)\sin(2\pi y)\;,\,\sin^2(\pi y)\sin(2\pi x)\right) \label{eq:vortex}
	\end{align}

\section{Conclusion}
The objective of this work was to improve the accuracy of two-phase flow simulations via widely-used software packages (e.g., OpenFOAM) that solve a scalar transport equation of the form given by Eq. \ref{eq:cmptrans} to capture fluid-fluid interfaces. While a full discretization of Eq. \ref{eq:cmptrans} was presented in Section \ref{sec:disc}, at the very minimum, the adoption of Eq. \ref{eq:angcmp} instead of the commonly used $\Lambda_f=1$ in Eq. \ref{eq:cnstcmp} should result in substantial improvements in accuracy. Corresponding changes in implementation are truly minor. While comparisons to other methods in this category are provided in \ref{appA}, a comparison against more established geometric and algebraic interface capturing techniques (e.g., PLIC \cite{debar1974PLIC,youngs1982PLIC}, CICSAM \cite{ubbink1999CICSAM}, THINC \cite{xie2017THINC}) is needed and postponed for future study.

\section*{Acknowledgements}
\label{sec:Acknow}
Funding was provided by the Stanford University Petroleum Research Institute (SUPRI-B) affiliates.  The Office of Basic Energy Sciences Energy Frontier Research Center under Contract number DE-AC02-05CH11231 is also acknowledged for financial support. Kirill Terekhov is acknowledged for discussions on Godunov schemes.

\appendix

\section{Comparison against refs. \cite{lee2015adaptive} and \cite{piro2013adaptive}}
\label{appA}

\begin{figure} [h] \vspace*{+2cm}
	\centerline{\includegraphics[scale=0.5,trim={0.3cm 8.6cm 0.2cm 3cm},clip]{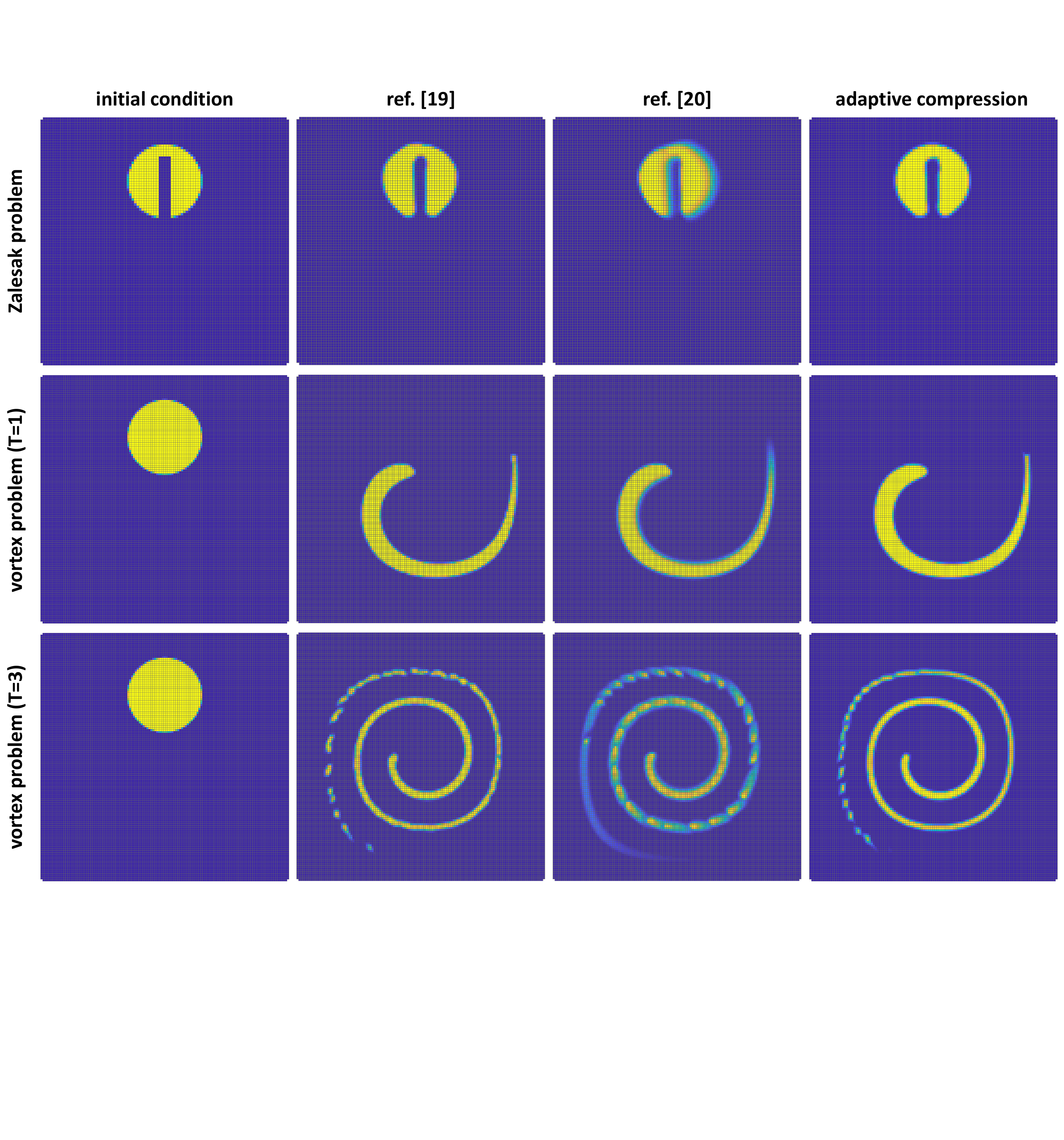}}
	\caption{Comparison between the adaptive compression of this work (Eq. \ref{eq:angcmp}) against the approaches proposed by \cite{lee2015adaptive} and \cite{piro2013adaptive}. In implementing \cite{lee2015adaptive} and \cite{piro2013adaptive}, we have simply replaced Eq. \ref{eq:angcmp} with the corresponding expressions for $\Lambda_f$ therein and have kept all spatial discretizations identical to that described in Section \ref{sec:disc}.}
	\centering
	\label{fig:appA}
	\end{figure}

Fig. \ref{fig:appA} compares the adaptive compression scheme of Section \ref{sec:disc} (Eq. \ref{eq:angcmp}) against approaches proposed by \cite{lee2015adaptive} and \cite{piro2013adaptive}. We consider tests 1-2 in Section \ref{sec:result}. In implementing \cite{lee2015adaptive} and \cite{piro2013adaptive}, we have simply replaced Eq. \ref{eq:angcmp} with the expressions for $\Lambda_f$ therein and have kept all spatial discretizations identical to that of Section \ref{sec:disc}; which are likely different from \cite{lee2015adaptive} and \cite{piro2013adaptive}. Fig. \ref{fig:appA} shows that Eq. \ref{eq:angcmp} results in noticeable improvements in capturing interfaces.
	
\bibliographystyle{elsarticle-num} 
\bibliography{References}

\end{document}